\documentclass[10pt]{article}
\usepackage[utf8]{inputenc}
\usepackage[T1]{fontenc}
\usepackage{dsfont}
\usepackage{url}
\usepackage{graphicx}
\usepackage{amsmath}
\usepackage[margin=1in]{geometry}
\usepackage{url}

\title{Photons guided by axons may enable backpropagation-based learning in the brain}

\author{Parisa Zarkeshian$^{1,2,3,5}$ \and Taylor Kergan$^1$ \and Roohollah Ghobadi$^{1,2,3}$ \and Wilten Nicola$^{4,1,3}$ \and Christoph Simon$^{1,2,3}$}

\date{
	\small{
	$^1$ Department of Physics \& Astronomy, University of Calgary, Calgary, AB, Canada \\
	$^2$ Institute for Quantum Science and Technology, University of Calgary, Calgary, AB, Canada \\
	$^3$ Hotchkiss Brain Institute, University of Calgary, Calgary, AB, Canada \\
	$^4$ Department of Cell Biology and Anatomy, University of Calgary, Cumming School of Medicine, Calgary, AB, Canada \\
	$^5$ 1QB Information Technologies (1QBit), Vancouver, BC, Canada 
}}

\begin{document}

\flushbottom
\maketitle

\begin{abstract}
	\noindent Despite great advances in explaining synaptic plasticity and neuron function, a complete understanding of the brain’s learning algorithms is still missing. 
	Artificial neural networks provide a powerful learning paradigm through the backpropagation algorithm which modifies synaptic weights by using feedback connections.  
	Backpropagation requires extensive communication of information back through the layers of a network. This has been argued to be biologically implausible and it is not clear whether backpropagation can be realized in the brain. 
	Here we suggest that biophotons guided by axons provide a potential channel for backward transmission of information in the brain.
	Biophotons have been experimentally shown to be produced in the brain, yet their purpose is not understood.
	We propose that biophotons can propagate from each post-synaptic neuron to its pre-synaptic one
	to carry the required information backward. 
	To reflect the stochastic character of biophoton emissions, our model includes the stochastic backward transmission of teaching signals.
	We demonstrate that a three-layered network of neurons can learn the MNIST handwritten digit classification task using our proposed backpropagation-like algorithm with stochastic photonic feedback. 
	We model realistic restrictions and show that our system still learns the task for low rates of biophoton emission, information-limited (one bit per photon) backward transmission, and in the presence of noise photons.
	Our results suggest a new functionality for biophotons and provide an alternate mechanism for backward transmission in the brain.  
\end{abstract}

\section*{Introduction}	
 
Learning is the process of gaining or improving knowledge or behavior by observing or interacting with the environment \cite{marton2013learning,gross2015psychology,rogers2010teaching}. 
In the brain, learning is dependent on the synaptic modifications between neurons \cite{hebb2005organization,markram1997regulation,bi1998synaptic,payeur2021burst}. However, the realization of learning in the brain is not completely understood \cite{humeau2019next}.
Existing theories mainly focus on neural
electrochemical signals and study their capabilities to be the brain’s information carriers \cite{dayan2001theoretical}. 

Backpropagation is an important part of our current understanding of learning in artificial neural networks and is most often used to train deep neural networks \cite{goodfellow2016deep}. Inspired by the fact that the brain learns by modifying the synaptic connections between neurons, the error signals are fed back to inner layers to update synaptic weights \cite{rumelhart1986learning,hecht1992theory}.
The broad applications and success of backpropagation and backpropagation-like algorithms as well as its core idea of using feedback connections to adjust synapses encouraged us to investigate if the brain's learning process is based on the principle of the backward flow of information \cite{zipser1988back,lillicrap2013preference,cadieu2014deep,khaligh2014deep,lillicrap2020backpropagation,payeur2021burst,sacramento2018dendritic,theories}.
However, it is not clear if and how backpropagation is implemented by the brain. It has been argued that some of its main assumptions such as having exactly the same weight for each feedback connection and its feedforward counterpart as well as the need for separate distinct forward and backward pathways of information are biologically unrealistic \cite{guerguiev2017towards}. 
Recent works suggest that symmetric weights are not necessary for effective learning \cite{kovsvcak2010stochastic,lillicrap2016random,lee2015joint,liao2016important,samadi2017deep,moskovitz2019feedback}; however, they are implicitly assuming a separate feedback pathway  \cite{guerguiev2017towards,lillicrap2020backpropagation}.
In this paper, we suggest a new potential photonic mechanism for the backward flow of information that avoids the above mentioned assumptions.

Biophotons are spontaneously emitted by living cells in the range of near-IR to near-UV frequency (350 nm–1300 nm wavelength) with low rates and low intensity, on the order of $1-10^3$ photons/(s.cm$^{2}$) \cite{cifra2014ultra}. These photons have been observed from microorganisms including yeast cells and bacteria \cite{konev1966very,vogel1999weak}, plants and animals \cite{prasad2013towards}, and different biological tissues \cite{kobayashi2009imaging,prasad2011two} including brain slices \cite{kobayashi1999vivo,tang2014,wang2011spontaneous}, yet it is unknown whether they have a biological function. In 1999, Kobayashi et al.\, performed in vivo imaging of biophotons from a rat's brain for the first time \cite{kobayashi1999vivo}. They demonstrated the correlation between biophoton emission intensity and neuronal activities of the brain with electroencephalographic techniques and suggested that biophoton emission from the brain originates from mitochondrial activities through the production of reactive oxygen species \cite{kobayashi1999vivo}. Moreover, several experiments studied the response of neurons and generally the brain to the external light \cite{leszkiewicz,wade1988mammalian,vandewalle2009light, starck2012stimulating,zhang2020violet} and showed that the brain has photosensitive properties.

The existence of these biophotons as well as the evidence that opsin molecules deep in the brain respond to light \cite{zhang2020violet} prompt the question of whether biophotons could serve as communication signals guided through the brain\cite{zarkeshian2018there}.
Axons have been proposed to be potential photonic waveguides for such optical communication \cite{kumar, sun2010biophotons, zangari2018node}.
The detailed theoretical modeling of myelinated axons shows that optical propagation is possible in either direction along the axon \cite{kumar}. 
Recent experimental evidence for light guidance by the myelin sheath supports the theoretical model \cite{depaoli2020anisotropic}.
Also, there is some older indirect experimental evidence in supporting light conduction by axons \cite{tang2014,sun2010biophotons}. 
Given the advantage that optical communication provides in terms of precision and speed in a technical context and the growing evidence that photons are practical carriers of information, one may wonder whether biological systems also exploit this modality.

If any backpropagation-like algorithm is employed by the brain, biophotons guided through axons
are a plausible choice for carrying the backward information in the brain in addition to the well-known electrochemical feed-forward signaling. 
\begin{figure}[t!]
	\centering
	\includegraphics[width=0.7\textwidth]{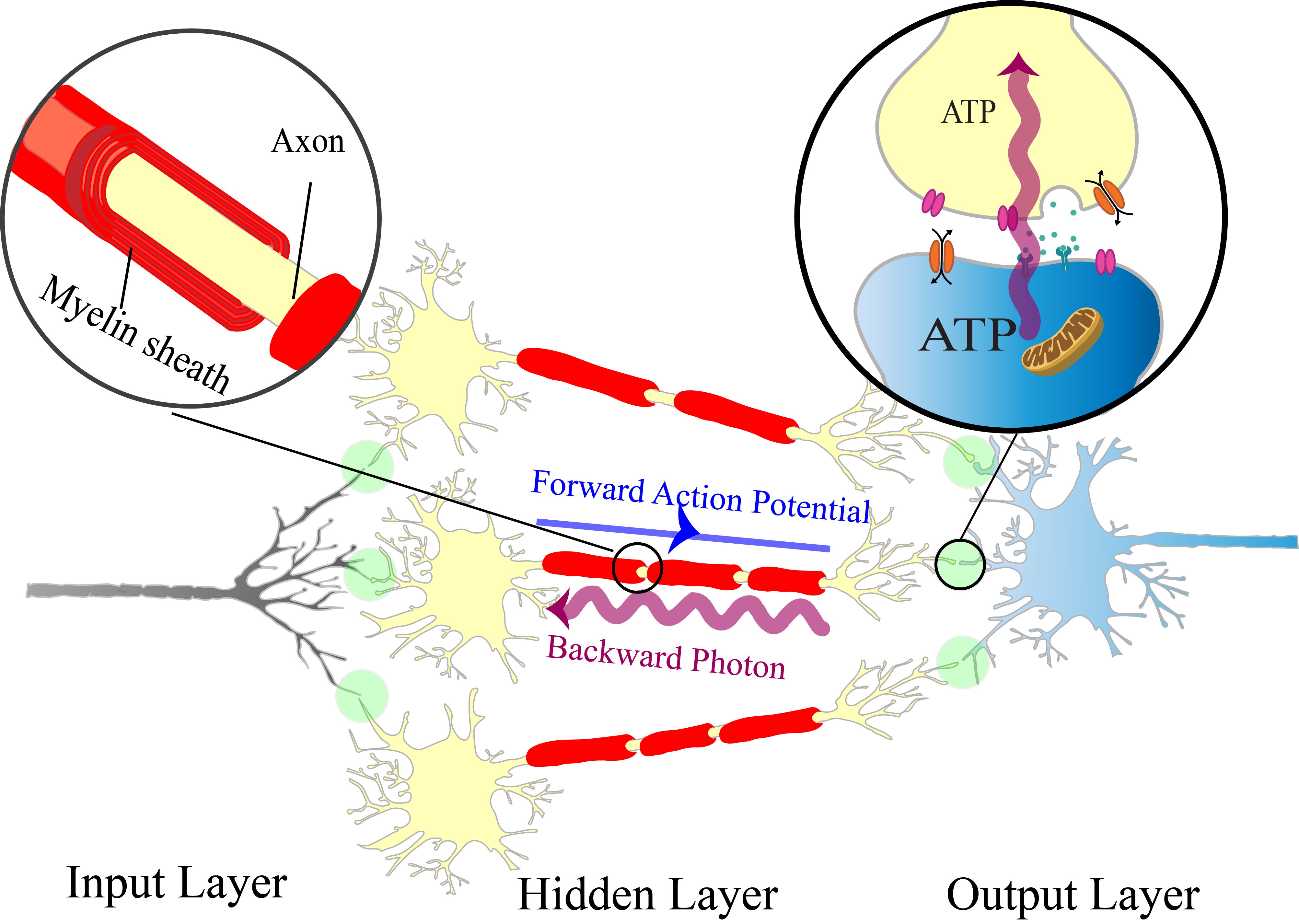}
	\caption[Schematic of a network, and distribution of ATP consumption]{Schematic of a simplified network  of neurons trained by backpropagation with stochastic photonic feedback. Three sets of neurons are represented as input, hidden, and output layers. For clarity, three neurons are shown in the hidden layer here. Connection of the dendrite of the post-synaptic neuron blue and the axonal terminal of the pre-synaptic neuron yellow at the synaptic cleft is enlarged (top-right).
	The strength of the synapse (or synaptic weight) is greater as the result of more working ion channels (orange oval-shaped gates) in the post-synaptic neuron \cite{voglis2006role}. This is in accordance with the greater amount of ATP (adenosine triphosphate, the energy carrier molecules) usage in the post-synaptic neuron~\cite{harris2012}. That results in more biophoton production by the post-synaptic mitochondrion which can transfer backward information to the pre-synaptic neuron.
	The myelinated axon (top-left, enlarged) can guide the received backward photons along the axon.
	}
	\label{Fig1}
\end{figure}	 
We model the backward path of information as a communication channel (see Fig~\ref{Fig1}) in which photons are produced stochastically with fairly low rates as it is expected by experimental observations of biophotons from the brain \cite{kobayashi1999vivo,tang2014}.
The stochastically emitted biophotons update a random subset of synaptic weights in each training trial, meaning that only a percentage of the neurons transmit backward information at any given time.
We consider realistic conditions and evaluate the learning efficacy of the mechanism.
We demonstrate that even with a small proportion (a few percent) of neurons sending stochastic biophotons backward to the upstream neuron, networks with one hidden layer and photonic emission can still learn a complex task.
We examine our model for the case that photons carry only one bit of information.
We further incorporate noise (e.g.\ due to ambient light) in our model. 
Our results show that the network can still learn the task of MNIST digit recognition considering these realistic imperfections.

\section*{Results}
\begin{figure}[t!]
	\centering
	\includegraphics[width=\textwidth]{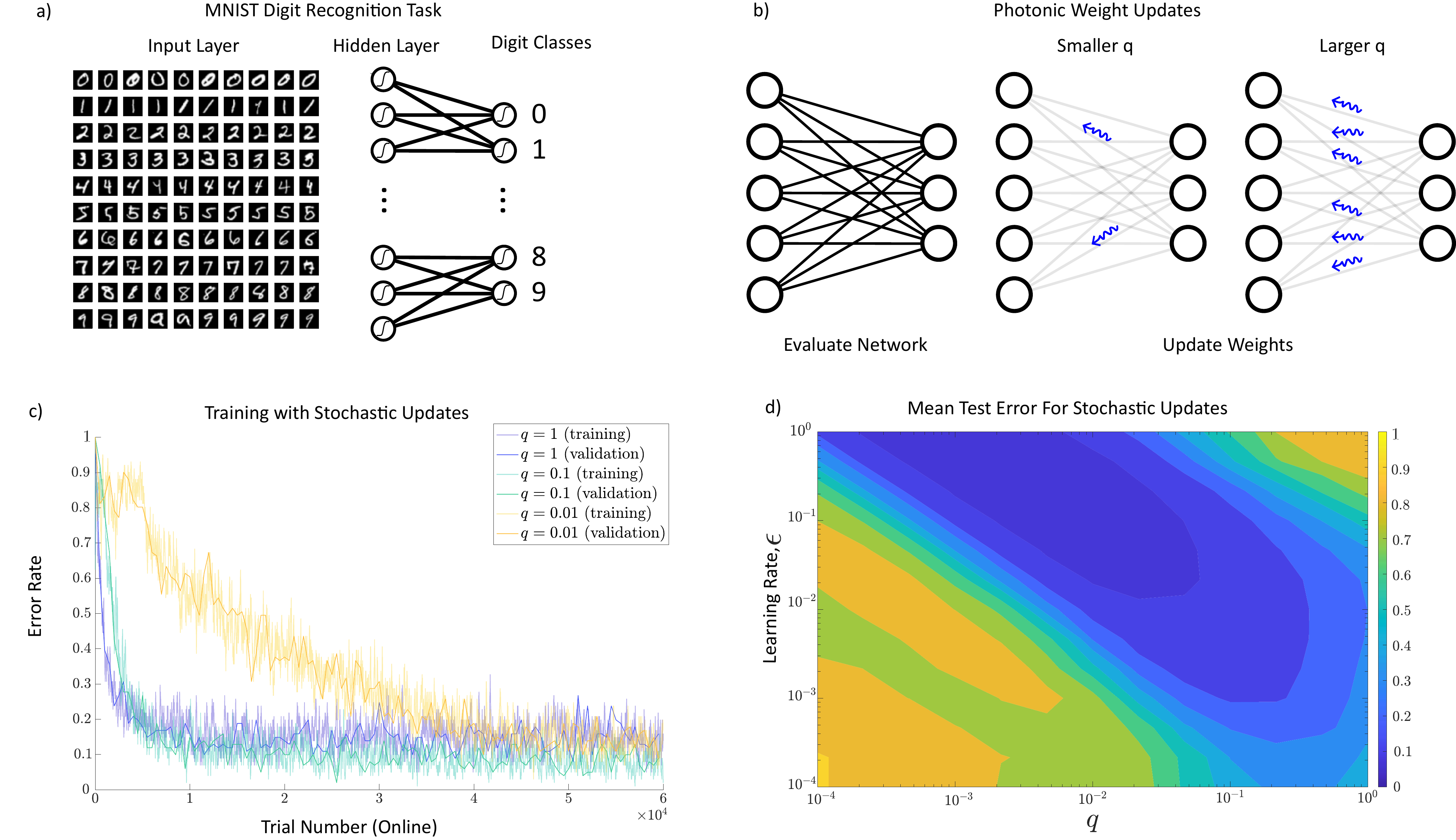}
	\caption[Training of a 3-layered artificial neural network with 500 neurons in the hidden layer with the stochastic photonic updates]{Training of a 3-layered artificial neural network with 500 neurons in the hidden layer by the stochastic photonic updates. \textbf{a,} The network is trained for an MNIST hand-written recognition task. For each training trial, the network receives sets of handwritten digits as input and their corresponding digit classes as the target output. 
	\textbf{b,} The network computes the feedforward weights and updates them with stochastic photonic feedback. The parameter \textit{q} is the probability of transmitting one photon per neuron, and as it gets larger (closer to 1), the network sends more backward photons and behaves closer to the conventional backpropagation algorithm. \textbf{c,} As the trial number grows, the error rate (that is the moving average of the past 100 trial errors, see Eq~\eqref{eq:error_train_n}) converges to a small value and the training completes. This convergence happens even for small values of \textit{q} but after greater numbers of trials. Here, the learning rate, $\epsilon=0.01$, has been kept small for the stability of the network. 
	\textbf{d,} The test error, which measures the distance between the target and the output of the trained model, is averaged over 10 repetitions of the test experiment for each different values of \textit{q} and $\epsilon$ (see Methods for details.)}
	\label{Fig2}
\end{figure}
In an artificial neural network, the backpropagation learning algorithm  calculates the gradient of an error function for each individual synapse with respect to the network weights and propagates the gradients backward all through the network to the upstream neuron. 
The forward flow of information is due to action potentials and action potentials go one way through the neural paths \cite{purves2008neuroscience}.
Our suggested mechanism for backward communication that determines the error signals does not interfere with the forward flow of information, as the electrochemical signaling pathway is  not likely manipulated by biophotons on short time scales. 
We do not require a separate network of neurons for feedback, addressing one of the main biologically problematic assumptions of backpropagation.

In our experiments, we consider a network with three sets of layers that are categorized into three classes input, hidden, and output layers (Fig~\ref{Fig1}). The hidden layer consists of 500 units (neurons) and the number of neurons in the other two layers depends on the task to be trained. The goal of the training is to reduce the loss function (Eq~\eqref{err}) which is the distance between the target output and the calculated output of the network. The synaptic weights are updated stochastically to mimic the random emission and propagation of biophotons that carry the information backward. 
We train the network for the 
MNIST digit recognition task \cite{lecun1998mnist,grother1995nist} in an online fashion. 
The mathematical details of the model are described in the Methods section. 

\subsection*{Neurons are trained with stochastically backpropagated photons.} 
Here we provide numerical evidence that our described model  is trainable, even with partial backpropagation 
of errors (teaching signals), 
by testing it on
the classification task of MNIST digits. The MNIST dataset of handwritten digits consists of 60,000 training examples and 10,000 test examples. Each training example in the dataset is a grey-scale image of 28 by 28 pixels of handwritten single digits between 0 and 9, in which the handwritten digits are recognized. The task is to classify a given image of a handwritten digit into one of the  10 classes (see Fig~\ref{Fig2}a). After evaluating the network and calculating the weights of connections, a random proportion $q$ of the neurons releases photons. The photon travels backward to transmit the error signal to its pre-synaptic neuron (see Fig~\ref{Fig1}) to update the pre-synaptic weights  (see Equations~\eqref{wt1}-\eqref{wt2} and the discussion for Stochastic photonic updates afterward in Methods). For small values of $q$, the photon emission is sparser and as it gets closer to 1, more photons travel backward (see Fig~\ref{Fig2}b), and the model performs closer to the original backpropagation algorithm where all weights get updated (see Methods, Stochastic photonic updates discussion). For stability, we keep the learning rate $\epsilon$  small (the learning rate is a tunable parameter between 0 and 1 that controls the step size of each iteration of weight updates in the training of the neural network) and of the order of $0.001$ in most of the simulations. We show that training error converges to a small constant value  after  $6\times 10^4$ number of trials for reasonable values of $q$. Here, the training error is the average error of the last 100 samples. While the training is in progress, we use a validation dataset  
to estimate how accurately the model performs and avoid overfitting. The validation error is computed every 500 steps. Fig~\ref{Fig2}c shows that we can still train a network for small values of $q$ of the order of $10^{-4}$. This shows that with even low emission rates of biophotons, the backpropagation-like channel can still learn the task. 
We also compare the output of the trained model with the expected output (test dataset, independent from the training dataset) and calculate the test error to check the performance of the trained model, shown in Fig~\ref{Fig2}d. 
Next, we restrict the amount of information that each photon transmits back in the model.

\begin{figure}[ht!]
	\centering
	\includegraphics[width=\textwidth]{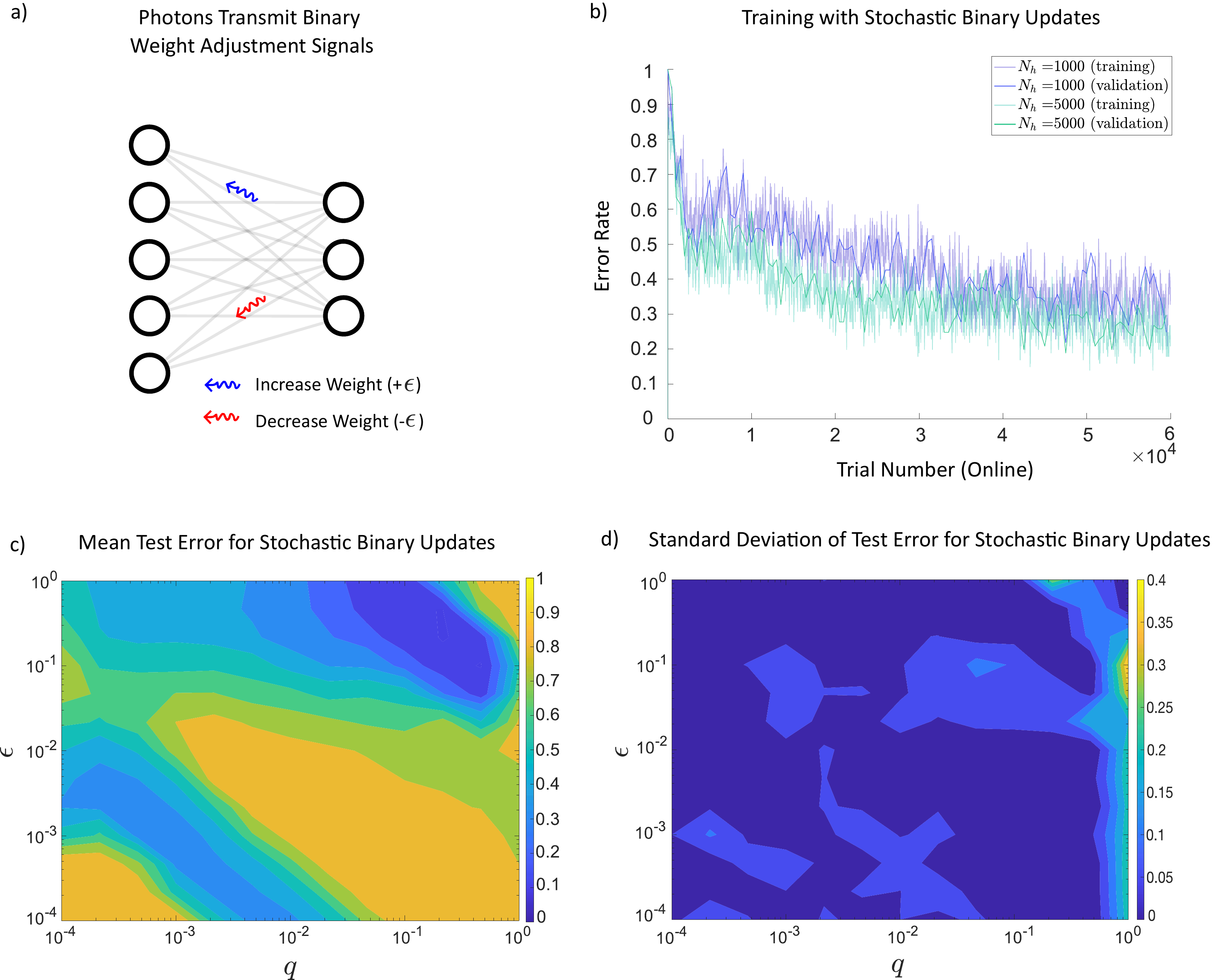}
	\caption[Training of a 3-layered artificial neural network with stochastic photonic updates and binary signal limit]{Training of a 3-layered artificial neural network by the stochastic photonic updates with carrying one bit of information. \textbf{a,} Stochastic weight updates transmit binary information and each update either increases or decreases the weight by a fixed amount, $\epsilon$. For more details see Eq~\eqref{wt1_sgn} and Eq~\eqref{wt2_sgn} in Methods. \textbf{b,}
	Here, $q=0.1$ and $\epsilon=0.1$. With stochastic binary updates, the training of MNIST classification task is still successful. As the number of hidden units increases the error rate converges to smaller values. \textbf{c,} To obtain mean test error, for each values of $\epsilon$ and $q$, 10 different networks with 500 hidden units were evaluated and the test error was averaged. \textbf{d,} Standard deviation of the test error is calculated per mesh point.}
	\label{Fig3}
\end{figure}

\subsection*{Neurons can learn even if each stochastic backpropagated photon carries only one bit of information.}\label{sec:sign}

As it may not be realistic to assume that a single photon can carry unlimited detailed information, we investigated if photons with more limited information could still lead to learning. In order to limit the amount of information carried by each photon, we discretize the gradient information into binary increases ($+\epsilon$) and decreases ($-\epsilon$) as shown in Fig~\ref{Fig3}a with blue and red signals, see Methods Eq~\eqref{wt1_sgn}-\ref{wt2_sgn} for the implementation of this specific weight update.
Depending on their type (e.g.\ two different polarizations or frequencies), they might increase or decrease the weight by a fixed amount of $\epsilon$.
Fig~\ref{Fig3} summarizes the result of implementing this limitation of information transmission and that the network can still learn. In Fig~\ref{Fig3}b, we have increased the size of the network from 500 units in the hidden layer to 1000 and 5000 units. As we increase the size of the network, the error rate converges to smaller amounts. In Fig~\ref{Fig3}c, the test error demonstrated with respect to each mesh point has been averaged over 10 trials of training the different networks. The standard deviation of the trials is shown in Fig~\ref{Fig3}d.

\subsection*{The network can be trained even in presence of uncorrelated random photonic updates.}

\begin{figure}[ht!]
	\centering
	\includegraphics[width=\textwidth]{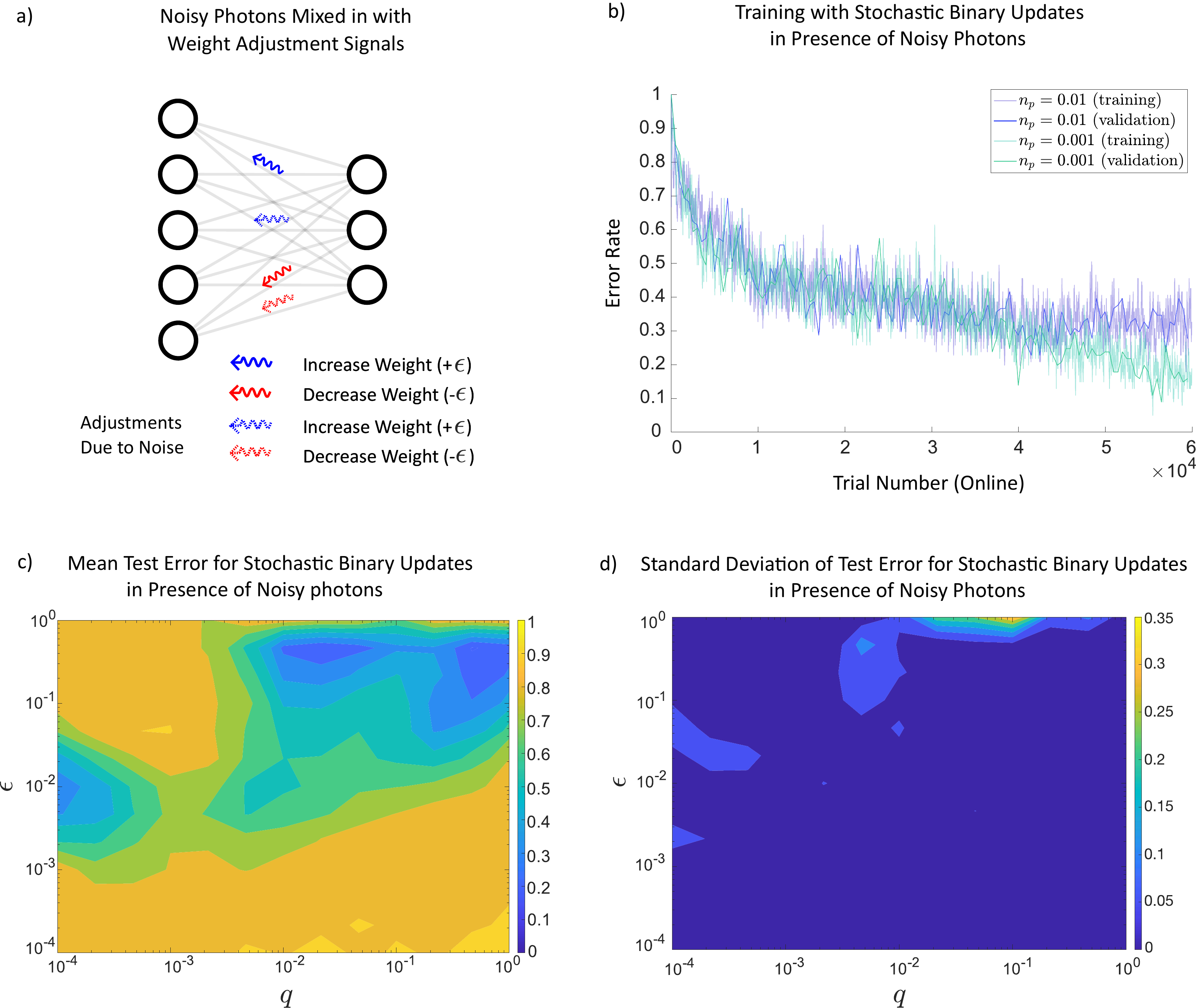}
	\caption[Training of a 3-layered artificial neural network with the noisy stochastic photonic updates and binary signal limit]{Training of a 3-layered artificial neural network by the noisy stochastic photonic updates with binary signal limit. \textbf{a,} After evaluation of the network, binary photonic signals are transmitted back to update weights, however, in the presence of uncorrelated noisy photons. \textbf{b,} Here, the deterministic photon emission rate, $q=0.1$, and the learning rate, $\epsilon=0.1$, are given for a network with 500 units in the hidden layer. The error rate depends on the noisy photon rate, $n_p$ which defines the probability of emitting noisy photon update. \textbf{c,} The average error rate after 10 trials has increased due to the noise in the system. Here, $n_p=0.01$. There are still areas in the graph where learning is happening. This figure is generated after training 10 different networks for each mesh point and taking the average on the test error. \textbf{d}) The standard deviation of the test error after running 10 different networks.}
	\label{Fig4}
\end{figure}

As the biophoton emission rates are low, the impact of ambient light as noise should be considered \cite{kuvcera2013cell}. Although waveguiding of biophotons by the axons would mitigate the effect of such noise \cite{kumar}, it is still important to consider. Other possible sources of noise could be some uncorrelated photon emissions and dark counts by detectors. 
We investigated biophoton-assisted learning in the presence of noise. We model the noise as uncorrelated random photons that impair the photonic updates. 
As shown in Fig~\ref{Fig4}a, noise photons (dashed ones) are also backpropagated. They disturb the process of weight adjustment and increase the error rate (see Fig~\ref{Fig4}b). The model was simulated for a 3-layered network with 500 units in the hidden layer and $q=0.1$ for different values of $n_p$, where $n_p$ is the proportion of neurons that emit noise (meaningless) photons. As long as the $n_p$ is smaller than $q$ (signal to noise ratio is smaller or equal to one), the training error converges according to Fig~\ref{Fig4}b. 
The comparison between Fig~\ref{Fig3}c and Fig~\ref{Fig4}c shows that for some areas of the parameter space learning still works even in the presence of noise with very low standard deviation of test error (see Fig~\ref{Fig4}d).

\section*{Discussion}

We have shown that backpropagation-like learning is possible with  stochastic photonic feedback, inspired by the idea that axons can serve as photonic waveguides, and taking into account the stochastic nature of biophoton emission in the brain. Considering realistic imperfections in the biophoton emission, we trained the network when each photon carried only one bit of information and showed that the network learned. We also examined the learning in presence of background noise and our results demonstrated its success. Here we discuss the biological inspiration for our suggested mechanism, address a few related questions, and propose experiments to test our hypothesis.

Synaptic weights are considered as the amount of influence that firing a pre-synaptic neuron has on the post-synaptic one \cite{hebb2005organization,markram1997regulation,FROHLICH201647}. This is directly related to the number of ion channels affected in the post-synaptic neuron \cite{debanne2003brain,meriney2019synaptic}. 
The greater the synaptic weight is, the more ion channels are working in the post-synaptic neuron, which requires more metabolic activity \cite{harris2012,voglis2006role}. 
That, in turn, escalates ATP usage resulting in more active post-synaptic mitochondria \cite{Stoler2021}. 
As mitochondria in the post-synaptic neuron work harder and consume more energy, more reactive oxygen species (ROS) are produced \cite{pospivsil2014role,turrens2003mitochondrial,murphy2009mitochondria,lambert2009reactive}. 
The emission of biophotons has been linked to the production of reactive species such as ROS and carbonyl  in mitochondria \cite{pospivsil2014role,kobayashi1999vivo,miyamoto2014singlet,pospivsil2019mechanism}. 
Thus, a higher production rate of ROS leads to a higher production rate of biophotons in the post-synaptic neuron. 
That directly relates the production of biophotons in the post-synaptic neuron to the synaptic weight changes. Proportionality of the photon emission to the weights is part of what is required for backpropagation (see Eq~\eqref{eq:wt_err2} in Methods). 
In addition, neurons may have evolved to encode the error signals in the photonic flux, e.g.\ by modulating biophoton emission as a function of incoming biophotons received.

An important question is how photonic information could be relayed across multiple network layers. 
Opsins are well-known for their ability of light detection in retina \cite{buhr2015neuropsin} and skin \cite{buhr2019neuropsin} of mammals. But they also exist in the deep brain tissues of mammals \cite{yamashita2014evolution} and are highly conserved over evolution. The existence of such light absorbent proteins in deep brain tissues suggests that they might serve as biophoton detectors. 
Moreover, a biological effect of external light mediated by opsins deep in the brain has recently been demonstrated, namely the opsin-mediated suppression of thermogenesis (heat production) in response to light \cite{zhang2020violet}.
On the other hand,  mitochondria always balance ATP production versus thermogenesis \cite{li2020mfsd7c}. 
Such suppression of thermogenesis via opsin-mediated photon detection could lead to more production of ATP by mitochondria which results in more photon production. Thus, it could constitute a relay across the neuron in the photonic backpropagation channel.

In our modeling, we have considered the fact that the amount of information carried by each photon may be limited, for example to one bit. This information could be encoded in the polarization of the photons, or in their wavelengths \cite{senior2009optical, hui2019introduction, nielsen_chuang_2010}. The amount of information that can be successfully transmitted also depends on the detection mechanism, e.g.\ there could be different opsins responding to different wavelengths.
 
Although low rates of biophoton emission might be a concern\cite{kuvcera2013cell}, guiding them by axons could be part of the solution because it will limit the loss of signal photons and reduce the impact of background light \cite{kumar}. Biophoton emission rate from a slice of a mouse brain was measured at one photon per neuron per minute \cite{tang2014} at rough estimation. This reported rate is one to two orders of magnitude lower than the electro-chemical signaling rate in the brain \cite{buzsaki2014log}. 
If biophotons are guided through the axons, it should be noted that the measured rates of brain emission only reflect the scattered photons and there could be more light propagating in a guided way than the experimental observations from the outside.

To verify the role of biophotons in learning in the brain, we propose some in-vivo experimental approaches. One type of tests is to genetically modify possible photon detectors in the brain, such as opsins, using well-studied optogenetics methods \cite{deisseroth2015optogenetics,adamantidis2014optogenetics,beyer2015optogenetic} in order to impair biophoton reception by the network and observe the effects on the learning process. Another type of test could be using the RNA interference process \cite{hannon2002rna,summerton2007morpholino,gao2021active} in non-genetically engineered animals to target the silencing of specific sequences in genes that involve the generation or reception of biophotons, which could affect learning.
Also, one could introduce background light into the neural network in vivo or add noise into the axon to see if that affects the learning. We have modeled noise by adding uncorrelated photons into the network. One could implement the idea of extra uncorrelated photons by introducing luciferase and luciferin (whose reaction produces bioluminescence without requiring an external light source) into the brain of the living animal by using optogenetic tools \cite{land2014optogenetic,park2020novel}.  

It has been suggested that biophotons in the brain could transmit not only classical but also quantum information \cite{kumar,simon2019can,smith2021radical}, however, this still requires experimental confirmation. In the context of the present work, which is focused on a potential role for photons in learning, the possibility of transmitting quantum information by biophotons could be connected to the field of quantum machine learning \cite{paparo2014quantum,crawford2016reinforcement,xia2021quantum}, which studies potential advantages of quantum approaches to learning.

If the brain's biophotons are involved in learning by transmitting information backward through the axons, then it would reveal a new feature of the brain and can answer some fundamental questions about the learning process. It is also worth noting that 
our stochastic backpropagation-like algorithm might be of interest beyond the biophotonic context and could have applications in other fields such as neuromorphic computing \cite{esser2015backpropagation,torrejon2017neuromorphic,markovic2020physics} and photonic reservoir computing\cite{paquot2012optoelectronic,duport2012all,tanaka2019recent,argyris2022photonic,davies2018loihi}.

\section*{Methods}

\subsection*{Network Equations}  
We consider a basic 3-layer artificial neural network, with $N_i$ input nodes, $N_h$ hidden layers, and $N_o$ output nodes. We label the output or activity of each node with the variable $a^\mu_k$, where $\mu = i,h,o$ stands for the input, hidden, and output layers, respectively. 
Neurons of the hidden and output layers perform some non-linearity, $\sigma(x)$, on their inputs.
We introduce non-linearity into the network with help of the logistic function which is a differentiable activation function $\sigma(x)=\frac{1}{1+e^{-x}}$
and has a convenient derivative of ${\frac {d\sigma(x)}{dx}}=\sigma(x)(1-\sigma(x))$. In an artificial neural network, the non-linear activation function produces a new representation of the original data that ultimately allows the non-linear decision boundaries for the network. The network equations then are
\begin{eqnarray}
a^i_j&=& x_j, \quad j=1,2,\ldots N_i, \label{eq1}\\
a^h_k &=& \sigma\left(\sum_{j=1}^{N_i} w^{h,i}_{k,j}a^i_j +b^h_k\right), \quad k = 1,2,\ldots N_h, \label{eq2}\\
a^o_l &=& \sigma\left(\sum_{k=1}^{N_h} w^{o,h}_{l,k} a^h_k\right),\quad l=1,2,\ldots N_o, \label{eq3}
\end{eqnarray}
where $b^h_k$ is a commonly included ``bias term''.

\subsection*{Standard Backpropagation}

Suppose we consider a finite sequence of inputs $\{x_{[1]},\ldots x_{[m]} \}$ with a matched sequence of outputs $\{y_{[1]},\ldots y_{[m]} \}$ as the training data set and we want to train the network such that the network output $a^0_{[n]}$ approximates the target output $y_{[n]}$, as $n$ grows. Note that subscript $[n]$ denotes the corresponding vector values at iteration $n=1,2,\ldots$. In the online learning approach, the backpropagation algorithm iteratively updates the weights $w^{h,i}, w^{o,h}$ to minimize the loss (error function) at each time. For each trial, the error function $L_n$ will be 
\begin{eqnarray}\nonumber
L_n &=& \frac{1}{2}\sum_{l=1}^{N_o} ((\delta_l^o)_{[n]})^2 \\ \nonumber
&=& \frac{1}{2}\sum_{l=1}^{N_o}\left((a^0_l)_{[n]} - (y_l)_{[n]}\right)^2. \label{err}
\end{eqnarray}

\noindent After each forward pass of information, the weights should be updated such that the network output gets closer to the target output. Thus, for the next training trial ($[n+1]$), the weights  $w^{h,i}$ and $w^{o,h}$ are updated as 
\begin{align}\label{wt0}
\left(w^{o,h}_{l,k}\right)_{[n+1]} &= \left(w^{o,h}_{l,k}\right)_{[n]} - \epsilon \frac{\partial L_n}{\partial w^{o,h}_{l,k}}\bigg|_{w^{o,h}_{l,k}= \left(w^{o,h}_{l,k}\right)_{[n]}},\\
\left(w^{h,i}_{k,j} \right)_{[n+1]} &= \left(w^{h,i}_{k,j} \right)_{[n]} - \epsilon \frac{\partial L_n}{\partial w^{h,i}_{k,j} }\bigg|_{w^{h,i}_{k,j} = \left(w^{h,i}_{k,j} \right)_{[n]}},
\end{align}
where $\epsilon$ is the learning rate. After evaluating the requisite derivatives, we have the following:
\begin{eqnarray}
\left(w^{o,h}_{l,k}\right)_{[n+1]} &=& \left(w^{o,h}_{l,k}\right)_{[n]} - \epsilon \cdot (\delta_l^o)_{[n]} \cdot (a_k^h)_{[n]}, \label{wt1}\\
\left(w^{h,i}_{k,j} \right)_{[n+1]} &=& \left(w^{h,i}_{k,j} \right)_{[n]} - \epsilon \cdot (\delta_l^h)_{[n]} \cdot (x_j)_{[n]}, \label{wt2}
\end{eqnarray} 
where $(\delta_l^o)_{[n]}$ denotes the error signal of the output layer and $(\delta_l^h)_{[n]}$ denotes the error signal of the hidden layer, that are given by:
\begin{align}\label{eq:wt_err1} 
(\delta_l^o)_{[n]} &= (a^0_l)_{[n]} - (y_l)_{[n]}, \\	
\label{eq:wt_err2}
(\delta_l^h)_{[n]} &=  \left( \sum_{l=1}^{N_o} (\delta_l^o)_{[n]}   \cdot w^{o,h}_{l,k} \right) \cdot \sigma'\left( (a_k^h)_{[n]} \right). 
\end{align}
In order to update the weights, the error signal $(\delta_l^o)_{[n]}$ is transmitted back to the hidden layer and $(\delta_l^h)_{[n]}$ is transmitted back to the input layer.

\subsection*{Training error rate and Test error}

To evaluate the performance of the training trials we calculate 
the error of each trial, which is a function of the error signal of the output layer, given by 
\begin{equation}\label{eq:error_train_n}
	e_{[n]} = 1 - \mathds{1}\left((\delta_l^o)_{[n]}\right),
\end{equation}
where 
\begin{equation*}
	\mathds{1}(\alpha) = \left\{\begin{array}{lc}
	0 & \text{if~} \alpha = 0 \\
	1 & \text{o.w.}
	\end{array} \right.
\end{equation*}
The training error simply indicates whether the network output matches the target data.   
The training error rate is calculated as the moving average of the training errors over the past 100 trials. 

If the training is successful the error rate converges to a negligible value. 
To make sure the network has truly leaned the task, 
we evaluate the performance of the network by using a new set of data called the test data set. 
The test error for each test experiment is the average number of times where the network output does not match the target output of the test data set.

\subsection*{Proposed photonic feedback}

We propose 
photonic backward propagation of error signals 
that is modeled under three main realistic limitations. 

\begin{enumerate}
	\item  \textbf{Stochastic photonic updates.} 
	To model the stochasticity of biophoton emissions in the brain, 
	in our proposed system, instead of updating all the weights, we only adjust $w^{h,i}$ for a random $q.(N_h N_i)$ number of weights, and $w^{o,h}$ for a random $q.(N_h N_o)$ number of weights where $q$ is the proportion of neurons that release photons. As $q$ gets larger (closer to 1), more photons are transmitted backward and for the case of $q=1$, it is the same original backpropagation algorithm.
	
	\item   \textbf{Stochastic photonic updates carrying only one bit of information.}
	In our model, when photons only transmit one bit of backward information, instead of Eq~\eqref{wt1} and Eq~\eqref{wt2}, the weight updates for the determined random number of weights, $q.(N_h N_i)$ or $q.(N_h N_o)$, are given by:
	\begin{eqnarray}
	\left(w^{o,h}_{l,k}\right)_{[n+1]} &=& \left(w^{o,h}_{l,k}\right)_{[n]} - \epsilon \cdot \mathsf{Sgn}\left( (\delta_l^o)_{[n]} \cdot (a_k^h)_{[n]} \right), \label{wt1_sgn}\\
	\left(w^{h,i}_{k,j} \right)_{[n+1]} &=& \left(w^{h,i}_{k,j} \right)_{[n]} - \epsilon \cdot \mathsf{Sgn}\left( \left( \sum_{l=1}^{N_o} (\delta_l^o)_{[n]}   \cdot w^{o,h}_{l,k} \right) \cdot \sigma'\left( (a_k^h)_{[n]} \right) \cdot (x_j)_{[n]} \right), \label{wt2_sgn}
	\end{eqnarray} 
	where $\mathsf{Sgn}$ is the sign function defined as:
	\begin{equation*}
	\mathsf{Sgn}(x) = \left\{ \begin{array}{cr}
	1 & \text{if~} x>0 \\
	0 & \text{if~} x=0 \\
	-1 & \text{if~} x<0
	\end{array} \right. .
	\end{equation*}
	
	\item \textbf{Stochastic photonic updates carrying only one bit of information in presence of noise.} To model the noise in feedback updates, the weights are first updated according to Eq~\eqref{wt1_sgn} and Eq~\eqref{wt2_sgn}. Then we select a random $np.(N_h N_i)$ number of $w^{h,i}$ weights, and a random $np.(N_h N_o)$ number of $w^{o,h}$ weights where $np$ is the proportion of neurons that release uncorrelated noise photons. The new weight updates are given by 
	\begin{eqnarray}
	\left(w^{o,h}_{l,k}\right)_{[n+1]} &=& \left(w^{o,h}_{l,k}\right)_{[n]} - \epsilon \cdot (\eta^o_{l,k})_{[n]}, \label{wt1_noisy}\\
	\left(w^{h,i}_{k,j} \right)_{[n+1]} &=& \left(w^{h,i}_{k,j} \right)_{[n]} - \epsilon \cdot (\eta^h_{k,j})_{[n]}, \label{wt2_noisy}
	\end{eqnarray} 
	where $\eta^o_{l,k}$ and $\eta^h_{k,j}$ are independent random variables takings values over $\{-1, +1\}$.  
\end{enumerate}

\section*{Acknowledgments}

This work was supported by the Natural
Sciences and Engineering Research Council (NSERC)
through its Discovery Grant program as well as the CREATE grant ’Quanta’. Partial funding was also provided by the Mitacs Accelerate program.
The authors would like to thank Sourabh Kumar, Daniel Oblak, and Rana Zibakhsh Shabgahi for useful discussions. 

\section*{Author contribution statement}

C.S. and W.N. conceived the project. 
The theoretical approach was developed by P.Z., R.G, C.S., and W.N. The numerical simulations were performed by P.Z. and T.K. with guidance from W.N. The paper was written by P.Z. with feedback from the other authors.

\section*{Availability of data and materials}

The datasets analyzed during the current study as well as our machine learning codes are available on GitHub via: \url{https://github.com/pzarkeshian/Photonic-backprop}.

\end{document}